\documentclass[aps,prl,twocolumn,tightenlines,nofootinbib,superscriptaddress,floatfix]{revtex4-2}
\usepackage[utf8]{inputenc}
\usepackage{graphicx}
\usepackage{placeins}
\usepackage{mathrsfs}
\usepackage{amsfonts}
\usepackage{setspace}
\usepackage{cellspace}
\usepackage{amsmath,amssymb,bm}
\usepackage[colorlinks=true,linkcolor=blue]{hyperref}
\usepackage{xcolor}
\usepackage{slashed}
\usepackage{hhline,multirow,tabularx}  %
\usepackage{dcolumn}    %
\usepackage{url}        %
\usepackage{braket}     %
\usepackage{makecell}
\usepackage{bbm}
\usepackage[normalem]{ulem}
\newcommand{\dd}{\text{d}}

\DeclareMathAlphabet{\mathdutchcal}{U}{dutchcal}{m}{n}
\SetMathAlphabet{\mathdutchcal}{bold}{U}{dutchcal}{b}{n}
\DeclareMathAlphabet{\mathdutchbcal}{U}{dutchcal}{b}{n}

\begin{document}

\title{Gluon GPDs from hard-exclusive electroproduction \\ of heavy quarkonium at NLO}
\author{Yuxun Guo}
\email{yguo5@bnl.gov, corresponding author}
\affiliation{RIKEN BNL Research Center and Physics Department, Brookhaven National Laboratory, Upton, NY 11973, USA}
\affiliation{Nuclear Science Division, Lawrence Berkeley National Laboratory, Berkeley, CA 94720, USA}

\author{Xiangdong Ji}
\email{xdji@sjtu.edu.cn}
\affiliation{Tsung-Dao Lee Institute and School of Physics and Astronomy, Shanghai Jiao Tong University, Shanghai, China}

\author{Yao Ji}
\email{yaoji@cuhk.edu.cn, corresponding author}
\affiliation{School of Science and Engineering, The Chinese University of Hong Kong, Shenzhen 518172, China}
\affiliation{Southern Center for Nuclear-Science Theory, Institute of Modern Physics, Chinese Academy of Sciences, Huizhou, Guangdong 516000, China}

\author{Feng Yuan}
\email{fyuan@lbl.gov}
\affiliation{Nuclear Science Division, Lawrence Berkeley National Laboratory, Berkeley, CA 94720, USA}

\begin{abstract}

Gluon tomography of the nucleon through hard-exclusive electroproduction of $J/\psi$ has been one of the key motivations for Electron-Ion Collider. However, the next-to-leading order (NLO) perturbative QCD calculations available in the literature were performed either in the collinear factorization limit where the heavy-quark mass is neglected, or in heavy-quark limit where the heavy quarkonia are approximately point-like. We present a full next-to-leading order collinear factorization for the hard exclusive production of heavy mesons that retains the meson mass as well as full light-cone distribution amplitude. We compare the full NLO result with the special limits and estimate the errors due to potentially-large corrections from $\alpha_s(M)$ and $M^2/Q^2$ at this order. 
\end{abstract}

\maketitle

\textit{Introduction}---Understanding the quark and gluon structure of the nucleon within quantum chromodynamics (QCD) is a central goal of modern nuclear physics. Of particular interest are the origins of the nucleon's mass and spin and their connection to the spatial and momentum distributions of quarks and gluons encoded in the generalized parton distributions (GPDs)~\cite{Ji:1994av,Ji:1996ek,Muller:1994ses}. Hard exclusive production processes provide primary experimental access to GPDs and are therefore crucial to explore this multidimensional structure~\cite{Ji:1998pc,Goeke:2001tz,Diehl:2003ny,Ji:2004gf,Belitsky:2005qn,Guo:2026swz}. Gluon GPDs are particularly elusive, and their precise determination will be a major objective of the future Electron--Ion Collider (EIC)~\cite{Accardi:2012qut,AbdulKhalek:2021gbh}. %
Exclusive productions of heavy vector mesons, such as $J/\psi$ and $\Upsilon$, are especially sensitive to the gluon GPDs~\cite{Hoodbhoy:1996zg,Koempel:2011rc,Ivanov:2004vd,Chen:2019uit,Flett:2021ghh} and constitute the flagship channel for precision gluon tomography at the EIC~\cite{Accardi:2012qut,AbdulKhalek:2021gbh}. Complementary efforts at Jefferson Lab (JLab) seek to constrain gluon GPDs and their moments through near-threshold $J/\psi$ photoproduction at lower energies~\cite{GlueX:2019mkq,Duran:2022xag,GlueX:2023pev,007:2026dow,Chatagnon:2026qsv,Tyson:2026gnd,Gryniuk:2016mpk,Hatta:2018ina,Hatta:2019lxo,Mamo:2019mka,Mamo:2021krl,Guo:2021ibg,Guo:2023pqw,Sun:2021pyw,Sun:2021gmi,Kharzeev:2021qkd}.

Theoretically, collinear factorization for deeply virtual meson production (DVMP) in lepton-nucleon collisions has been well established in the asymptotic regime $Q^2\gg M^2$, where the longitudinal amplitude factorizes into a perturbative hard coefficient, nucleon GPDs, and a meson distribution amplitude (DA)~\cite{Radyushkin:1996ru,Collins:1996fb}. At leading power, however, the meson mass $M$ is neglected, leaving potentially sizable mass corrections at experimentally accessible $Q^2$. By contrast, nonrelativistic QCD (NRQCD)~\cite{Bodwin:1994jh}, widely used for exclusive heavy-meson photo- and electroproduction~\cite{Hoodbhoy:1996zg,Koempel:2011rc,Ivanov:2004vd,Chen:2019uit,Flett:2021ghh}, retains the heavy-quark mass but describes the bound state through an expansion in the relative heavy-quark velocity $v$, assuming $v^2\ll 1$. Since both finite-mass and relativistic corrections can be substantial at the relevant EIC kinematics, particularly for exclusive $J/\psi$ production~\cite{Guo:2024wxy,Mantysaari:2021ryb}, a unified framework that treats these effects simultaneously and consistently will be essential for precision gluon tomography at the EIC.

In this paper, for the first time, we present a %
QCD factorization framework for hard exclusive production of heavy mesons that retains the meson mass and factorizes the amplitude into a short-distance coefficient, nucleon GPDs, and meson DAs. %
This formalism recovers the conventional collinear factorization and NRQCD factorization in their respective limits. It thereby provides a unifying framework for meson-mass effects in the collinear description and relativistic corrections in NRQCD. 

We will demonstrate this factorization with an explicit calculation at one-loop order (next-to-leading order in QCD coupling). The collinear divergence associated with the GPDs will be absorbed into the renormalization of the GPDs. Additional logarithmic dependence on heavy quark mass can be included by the renormalization of the vector meson DAs. From the one-loop results, we obtain the consistent evolution equations for the GPDs and DAs. These will help us to make reliable predictions for precise extraction of nucleon GPDs from future experiments. To illustrate the sizes of corrections to the previous formalisms, we will show some numeric results for the typical kinematics at the EIC. 

Our derivations %
confirm previous results on the connection between the light-cone distribution amplitudes and the NRQCD factorization of exclusive heavy quarkonium productions mainly in decay processes~\cite{Ma:2006hc,Jia:2008ep,Jia:2010fw,Bell:2008er}. %
The method developed below will extend these calculations to a complete factorization framework where systematic improvements from both higher order perturbative QCD corrections and relativistic corrections can be carried out accordingly.

\textit{Power counting and factorization framework---}We consider the exclusive meson electroproduction,
\begin{equation}
\gamma^*(q)+N(p)\to M(p_M)+N'(p')\,,
\label{eq:process}
\end{equation}
with $Q^2=-q^2>0$, $p_M^2=M^2$, $P=(p+p')/2$, $\Delta=p'-p$, and $t=\Delta^2$. We take $Q^2\gg\Lambda_{\mathrm{QCD}}^2$ at fixed $\xi$ (see below) and $M^2/Q^2$, retaining the meson and heavy-quark masses in the hard scattering while neglecting corrections suppressed by $\Lambda_{\mathrm{QCD}}^2/Q^2$, $|t|/Q^2$, and $M_N^2/Q^2$, where $M_N$ is the nucleon mass. In the following, we take the longitudinal meson production as an example to demonstrate the factorization argument of our formalism with explicit one-loop derivations. They can be directly compared to previous results from both NRQCD calculations~\cite{Ivanov:2004vd,Chen:2019uit,Flett:2021ghh} and conventional collinear factorization~\cite{Radyushkin:1996ru,Collins:1996fb}.

We carry out the derivations in a frame that $P_\perp=q_\perp=0$ and parameterize the virtual photon momentum as $q^\mu\simeq- Q^2/(2q^-)P^\mu+q^-n^\mu$, and the meson momentum $p_M^\mu\simeq M^2/(2q^-)P^\mu+q^-n^\mu$ with $q^-=(Q^2+M^2)/(4\xi)$ that preserves $q^2=-Q^2$ and $p_M^2=M^2$. 
Here, $n^2=0$ is a light-like vector with normalization $P\cdot n=1$. The skewness is hence defined as $\xi=-\Delta\cdot n/2$.
For the heavy-quark pair, we write the momenta as $p_Q\simeq z p_M+k_\perp$ and $p_{\bar Q}\simeq\bar z p_M-k_\perp$, where $k_\perp$ is the transverse momentum carried by the heavy quark respect to the quarkonium moving direction, $\bar z=1-z$ and $m_Q$ denotes the heavy-quark mass. %
In the calculation, we retain the leading-power momentum components and the explicit dependence on $m_Q$, while neglecting power-suppressed contributions, such as those proportional to $k_\perp$ and their virtualities.

At the leading power in our framework, the virtual photon splits into a heavy quark/antiquark pair, which diffractively scatters off the nucleon target before it hadronizes into a longitudinal heavy meson. The scattering amplitude, from which the transition form factors are defined, can be schematically factorized into the GPDs, hard coefficients, and the associated distribution amplitude,
\begin{align}
\begin{split}
     {\cal A}_L(\xi,t)&={\cal N}_L\int_0^1 dz\,\Phi_\psi(z,\mu_\Phi)\\
     &\times\sum_i\int_{-1}^{1}\frac{dx}{x}C_{L,i}(x,\xi,z,m_Q;\mu's)F_i(x,\xi,t,\mu_H) \ ,\label{eq:factorization}
\end{split}
\end{align}
where $\Phi_\psi(z;\mu_\Phi)$ represents the heavy meson distribution amplitude that depends on the factorization scale $\mu_\Phi$ and $F_i(x,\xi,t;\mu_H)$
for the GPDs of flavor $i=u,d,s,c,g$ that depend on the factorization scale $\mu_H$. ${\cal N}_L$ is an overall kinematic factor of the longitudinal meson production process. The hard coefficients can be perturbatively expanded as $C_{L,i}=a_s C_{L,i}^{(0)}+a_s^2 C_{L,i}^{(1)}+\cdots$ with $a_s\equiv\alpha_s(\mu_R)/(4\pi)$. 

For vector mesons, both $H$ and $E$ GPDs~\cite{Ji:1996ek} enter and their amplitudes admit the factorized form. Defining their contribution to the scattering amplitudes $\mathcal A_L$ as $\mathcal{H}$ and $\mathcal{E}$ respectively, the longitudinal differential cross section reads
\begin{equation}
\begin{aligned}
\frac{\dd\sigma_L}{\dd t}
&=\frac{1}{16\pi\lambda(W,Q,M_N)}
\biggl[(1-\xi^2)|\mathcal H|^2\\
&\quad-2\xi^2\operatorname{Re}(\mathcal H\mathcal E^*)
-\left(\xi^2+\frac{t}{4M_N^2}\right)|\mathcal E|^2\biggr],
\end{aligned}
\label{eq:cross-section}
\end{equation}
with $W^2\equiv(p+q)^2$ and $\lambda(W,Q,M_N)\equiv(W^2+Q^2-M_N^2)^2+4Q^2M_N^2$. The skewness can be approximated as $\xi\simeq(Q^2+M^2)/[2(W^2-M_N^2)+Q^2-M^2]$; see also Ref.~\cite{Flett:2025chf}.

Notably, the mass corrections already emerge at the tree-level in this framework as,
\begin{align}
 C_{L,g}^{(0)}(x,\xi,z;m_Q)
 =\frac{4\pi x^2\xi}{(x+\xi-i0)(x-\xi+i0)}\frac{1}{z\bar z+\kappa},
\label{eq:born}
\end{align}
with $\kappa\equiv m_Q^2/Q^2$, vanishing quark coefficient $C_{L,q}^{(0)}$ at LO, and ignoring the intrinsic heavy quark contributions. At $z=1/2$, $1/(z\bar z+\kappa)=4Q^2/(Q^2+4m_Q^2)$ reproduces the massive NRQCD coefficient, while $m_Q\to0$ at fixed $0<z<1$ gives the collinear weight $1/(z\bar z)$. This effect at the tree level was already addressed in Ref.~\cite{Guo:2024wxy}.

Beyond the tree level, the infrared (IR) divergence %
entirely comes from the collinear emissions of massless partons and can be absorbed into the renormalized nucleon GPDs. Then, the collinear emissions from heavy quarks, which are naturally regulated by the quark mass, will lead to logarithmic $m_Q$ dependences dictated by the evolution of heavy meson DA and heavy quark GPD. With this in mind, we develop the factorization framework using a two-step strategy and derive the factorization formula above with an explicit calculation at one-loop order in the following.

\begin{figure}[t]
 \centering
 \includegraphics[width=0.241\textwidth]{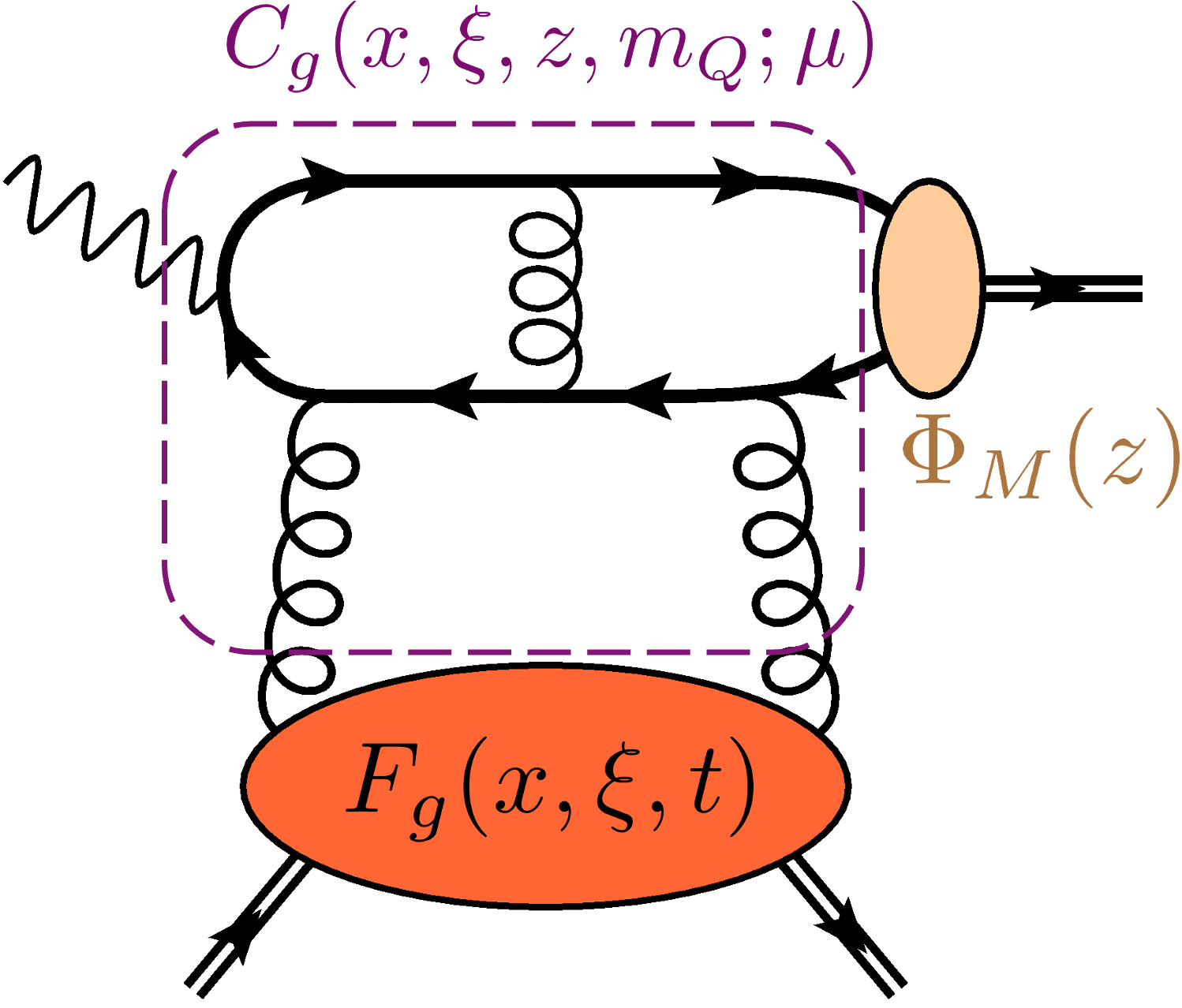}\hfill
 \includegraphics[width=0.241\textwidth]{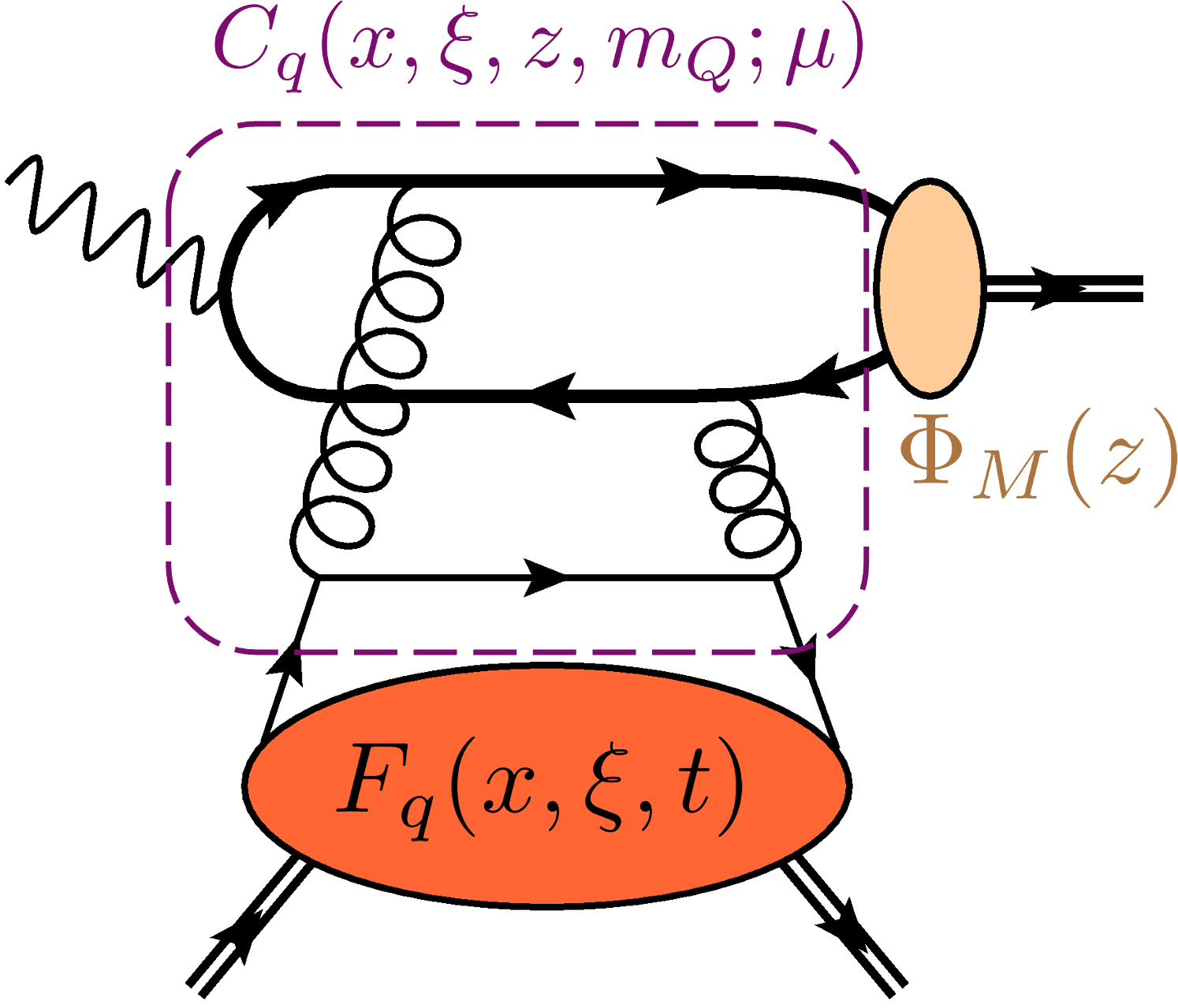}
\caption{An illustration of the collinear factorization in terms of the meson DA for heavy quarkonium productions with examples of NLO Feynman diagrams in the gluon (left) and light-quark (right) channels, assuming vanishing intrinsic heavy quark distribution. Heavy quarks are represented by thick fermion lines.}
 \label{fig:feyndiag}
\end{figure}

\textit{One-loop factorization with massive quark---}At one-loop order, we consider Feynman diagrams with all possible gluon radiation contributions, including hard, collinear and soft ones, as shown in Fig.~\ref{fig:feyndiag}. Dimensional regularization with $D=4-2\epsilon$ in $\overline{\mathrm{MS}}$ subtraction scheme will be applied to evaluate all these diagrams. First, as an important cross check, we find that the double poles in $1/\epsilon$, present in individual diagrams, indeed fully cancel. After UV renormalization, the remaining single poles have the universal form
\begin{equation}
\left.\widehat C_{L,i}^{(1),\mathrm{UV}}(m_Q)\right|_{\mathrm{IR}}
=-\frac1{\epsilon_{\mathrm{IR}}}
 C_{L,g}^{(0)}(m_Q)\otimes V_{gi}^{(1)},
\label{eq:dimreg-pole}
\end{equation}
where $i=g,q$, $V_{ij}^{(1)}$ stands for the one-loop GPD evolution kernel, and $\otimes$ represents the standard convolution of $x$ or $z$ for GPD/DA respectively. Then the counterterms from the nucleon GPD renormalization with a factorization scale $\mu_H$ remove the poles and yield IR finite coefficients. Nevertheless, they still contain collinear divergence regulated by $\ln(Q^2/m_Q^2)$ in the limit of $Q^2\gg m_Q^2$.

To study %
the collinear emissions of the heavy quark, we introduce its nucleon GPD and meson DA with factorization scales $\mu_{{H}}$ and $\mu_{\Phi}$ respectively as
\begin{align}
\!\!\!\!\!H_Q^+(\mu_{H})&=a_s\ln\frac{\mu_{H}^2}{m_Q^2}
 V_{qg}^{(1)}\otimes H_g(\mu_H)+O(a_s^2),
\label{eq:charm-matching}\\
\Phi_\psi(\mu_\Phi)
&=\Big[1+a_s\ln\frac{\mu_\Phi^2}{m_Q^2}
 V_\Phi^{(1)}\Big]\otimes\Phi_\psi(m_Q)+O(a_s^2),
\label{eq:da-evolution}
\end{align}
where $H_Q^+(x)=H_Q(x)-H_Q(-x)$ is the gluon-induced flavor singlet heavy-quark distribution and $V_\Phi^{(1)}$ is the one-loop Efremov--Radyushkin--Brodsky--Lepage (ERBL) evolution kernel of longitudinal DA~\cite{Efremov:1978rn,Lepage:1980fj}. For consistency, we use the same factorization scale $\mu_H$ for the heavy quark GPD. The intrinsic heavy-quark contribution has been excluded in the present work, which corresponds to a boundary condition $H_Q^+(\mu=m_Q)=0$ in Eq.~\eqref{eq:charm-matching} and can be included systematically. We will address this issue in a future publication.  %

By refactorizing the nucleon GPD and meson DA of the heavy quark in the hard coefficients with the evolved ones in Eqs. \eqref{eq:charm-matching} and \eqref{eq:da-evolution} with appropriate finite term subtractions based on boundary conditions, we obtain the coefficients $C_{L,i}$ that generally depend on $(x,\xi,z,m_Q;\mu_H,\mu_\Phi,\mu_R)$. %
This way, the scales $\mu_H$, $\mu_\Phi$ fully separate the collinear %
gluon radiations from the hard scattering, and all the logarithmic divergences in the $m_Q\to0$ limit cancel.
As expected, the hard coefficients $C_{L,i}$, still depend on $m_Q$ kinematically, and additionally on the renormalization and factorization scales. These scale dependencies follow the standard renormalization-group equations,  which at one loop reads:
\begin{subequations}
\label{eq:factorization-evolution}
\begin{align}
 \frac{\partial C_{L,i}^{(1)}}{\partial\ln\mu_H^2}
 &=-C_{L,j}^{(0)}\otimes_x V_{ji}^{(1)},
 \label{eq:wilson-gpd-scale-evolution}\\
 \frac{\partial C_{L,i}^{(1)}}{\partial\ln\mu_\Phi^2}
 &=-C_{L,i}^{(0)}\otimes_z V_\Phi^{(1)},
 \label{eq:wilson-meson-scale-evolution}\\
 \frac{\partial C_{L,i}^{(1)}}{\partial\ln\mu_R^2}
 &=\beta_0 C_{L,i}^{(0)}.
 \label{eq:wilson-renormalization-scale-evolution}
\end{align}
\end{subequations}
Here $i=g,q$, $C_{L,q}^{(0)}=0$, and LO beta function $\beta_0$ is defined from $d a_s/d\ln\mu_R^2=-\beta_0 a_s^2+O(a_s^3)$. Note that Eq.~\eqref{eq:wilson-gpd-scale-evolution} stands for the standard GPD evolution with quark-gluon mixing---while the intrinsic heavy quark is excluded, its mixing with the gluon appears in the hard coefficients at one-loop level.  The full expressions for the hard coefficients are too lengthy to present here and are provided in the ancillary files accompanying the arXiv submission in both \texttt{Mathematica} and \texttt{Python} formats.

\begin{figure*}[t]
 \centering
 \includegraphics[width=0.33\textwidth]{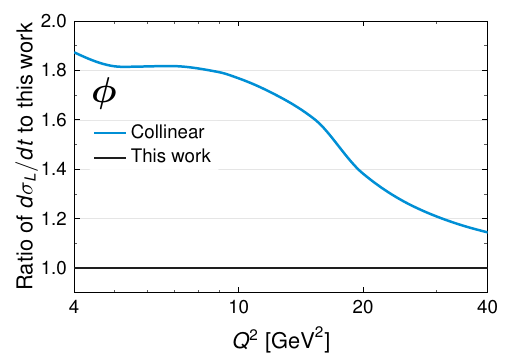}\hfill
 \includegraphics[width=0.33\textwidth]{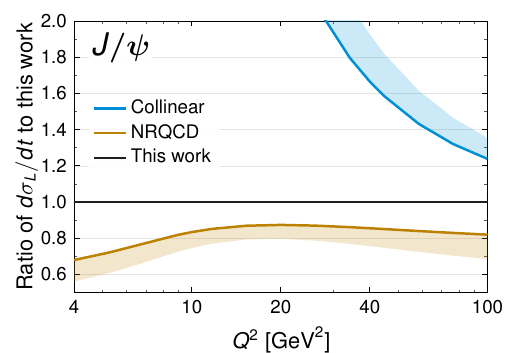}\hfill
 \includegraphics[width=0.33\textwidth]{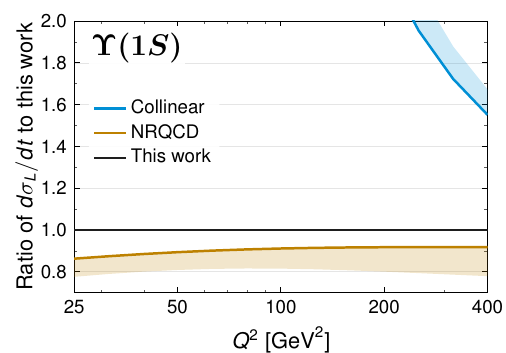}
\caption{Ratios of longitudinal differential cross sections for $\phi$ (left), $J/\psi$ (center), and $\Upsilon(1S)$ (right) at $W=50\,\mathrm{GeV}$ and $t=-0.1\,\mathrm{GeV}^2$, using NLO hard coefficients. Blue and amber curves denote collinear and NRQCD predictions divided by those of this work; black line marks unity. For quarkonia, solid curves use $c=0.5$ and shaded bands show the envelope of ratios at $c=0.5$ to $c=1$. Denominators use the corresponding $c$; the NRQCD numerator is independent of $c$ so that the amber bands are induced entirely by $c$-dependence from our predictions. The illustrative $\phi$ result uses $\varphi_\phi(z)=6z(1-z)$ and $m=M_\phi/2$.}

 \label{fig:xsec_benchmark}
\end{figure*}

Finally, we comment on the relation to the existing results in the literature. Since the $m_Q\to0$ limit is well-defined, the hard coefficients in this work are expected to reproduce the massless collinear results, which is indeed the case~\cite{Ivanov:2004zv,Duplancic:2016bge}. Meanwhile, the $z=1/2$ limit of this work also reproduces the NRQCD calculations~\footnote{We note a relative minus sign to Ref.~\cite{Flett:2021ghh} in the quark channel. Here we keep the sign of our current work.}. In particular, at the NRQCD limit, the perturbative correction to the DA can be separated out from our one-loop hard coefficients. Thus, the following DA with a perturbative tail extracted from $C_{L,i}^{(1)}$ provides an exact match to the NRQCD results at NLO,
\begin{equation}
\phi_\psi^{\mathrm{pert}}(z;\mu=m_Q)
=\delta(z-\tfrac12)+a_s C_F S_+(z)+O(a_s^2),
\label{eq:pert-da}
\end{equation}
where the dimensionless DA $\phi(z;\mu)$ has been defined according to $\Phi_\psi(z;\mu)=f_\psi \phi_\psi(z;\mu)$ and $S_+(z)$ is the one-loop perturbative tail calculated in Eq. (73) of Ref.~\cite{Wang:2013ywc} satisfying $\int_0^1dz\,S_+(z)=0$. Hence, this work reproduces both existing frameworks in appropriate limits, respectively. %

\textit{Phenomenology impact---}To quantify the phenomenological impact of this factorization framework, we consider the longitudinal electroproduction cross section of $J/\psi$ and $\Upsilon$ mesons at typical EIC kinematics of $W=50\,\mathrm{GeV}$ and $t=-0.1\,\mathrm{GeV}^2$, using the central unpolarized NLO GUMP fit as the GPD input~\cite{Guo:2025muf,Guo:2023ahv,Guo:2022upw} with necessary conversion since the original GUMP was developed with only two quark flavors. 

For heavy quarkonia we set $m_Q=M/2$ for simplicity and model the DA input at $\mu=m_Q$ by
\begin{equation}
\begin{aligned}
 \phi_{\mathrm{in}}(z;c)&=\phi_{\mathrm{np}}(z;c)+a_s C_F S_+(z),\\
 \phi_{\mathrm{np}}(z;c)&=N(c)z\bar z
 \exp\!\left[-\frac{(z-1/2)^2}{2\sigma_z^2}\right],
\end{aligned}
\label{eq:nonpert-da}
\end{equation}
with $\sigma_z=c\,\alpha_s(m_Q)$ and $N(c)$ the appropriate normalization. We choose $c=0.5$ as the %
typical value, %
which gives the moments $\langle(2z-1)^2\rangle_{\mathrm{np}}$ of about $0.092$ for $J/\psi$ and $0.039$ for $\Upsilon$. This is consistent with the leading-order nonrelativistic relation $\langle(2z-1)^2\rangle\simeq\langle v^2\rangle/3$ and the values $\langle v^2\rangle\simeq 0.3$ and $0.1$ estimated for $J/\psi$ and $\Upsilon$ with NRQCD, respectively, in Ref.~\cite{Grossman:2015cak}. We include the perturbative tail in the DA input so that the non-perturbative DA approaches the perturbative one in Eq.~\eqref{eq:pert-da} at the $c\to0$ limit.  To show the effects of the DA's width, we also vary $c$ from $0.5$ to $1.0$. 

The non-perturbative DA input (Eq.~\eqref{eq:nonpert-da}) with leading-logarithmic (LL) ERBL evolution will be used for the collinear calculation based on Eq.~\eqref{eq:factorization}, while the perturbative DA (Eq.~\eqref{eq:pert-da}) at fixed order enters the NRQCD calculation, so the effects of kinematic corrections as well as the relativistic resummation can be quantified.

The mid and right panels in Fig.~\ref{fig:xsec_benchmark} present the longitudinal electroproduction cross section of heavy quarkonia $J/\psi$ and $\Upsilon$ at the given kinematics in the three frameworks, respectively.  The central values are calculated at scales $\mu_H=Q$, $\mu_\Phi=\max(m_Q,Q)$, and $\mu_R=\max(2\,\mathrm{GeV},\sqrt{Q^2+M^2}/2)$. The GPDs are evolved to each varied $\mu_H$ with GUMP fitting uncertainties excluded. The intrinsic heavy quark GPDs are omitted so that the heavy quarks are only perturbatively generated through Eq.~\eqref{eq:charm-matching}. The comparison shows that the collinear framework substantially overestimates heavy-meson cross sections even at large $Q^2$, by approximately $67\%$ at $Q^2=40\,\mathrm{GeV}^2$ for $J/\psi$ and $127\%$ at $Q^2=200\,\mathrm{GeV}^2$ for $\Upsilon$, relative to the massive predictions. At lower $Q^2$, the discrepancy exceeds an order of magnitude, underscoring the importance of retaining meson-mass dependence. Much of this effect is captured by the midpoint Born-level ratio $(1+M^2/Q^2)^2$, as conjectured on phenomenological grounds in Ref.~\cite{Guo:2024wxy}. This also motivates further studies of the meson-mass effects in DVMP for lighter mesons. The left panel of Fig.~\ref{fig:xsec_benchmark} illustrates an extension to $\phi$ production using the asymptotic DA $\varphi_\phi(z)=6z(1-z)$ and an effective quark mass $m=M_\phi/2$, showing a significant suppression of cross-sections even at the medium to higher $Q^2\lesssim15$ GeV$^2$. Nevertheless, a systematic treatment of meson-mass and target-mass corrections for the lighter meson remains an open task beyond the scope of this work.

The comparison with the NRQCD framework also reveals substantial sensitivity to the non-perturbative structure encoded in the heavy-meson DA. As shown in Fig.~\ref{fig:xsec_benchmark}, the finite-width DA prediction with $c=0.5$ exceeds the leading-velocity NRQCD reference by up to $47\%$ for $J/\psi$ and up to $16\%$ for $\Upsilon$, demonstrating the phenomenological importance of going beyond the midpoint approximation. A central strength of our factorization framework is that the meson's nonperturbative structure is systematically encoded in a universal DA. Its shape can therefore be studied through NRQCD~\cite{Wang:2013ywc,Wang:2017bgv} or determined from lattice QCD~\cite{Ji:2013dva,  Blossier:2024wyx,LPC:2026vyv}. %
This separation of universal meson structure from perturbative hard scattering provides a systematic route to predictions beyond the leading nonrelativistic approximation.

\textit{Conclusion}---In summary, we have established a factorization framework for longitudinal heavy-vector-meson electroproduction in terms of the meson DA that retains the heavy-quark mass. After perturbative matching, the hard coefficients at one-loop order reproduce conventional collinear factorization and leading-velocity NRQCD in their respective limits, providing a common description of finite-mass effects and meson structure beyond the nonrelativistic approximation.

Our numerical results demonstrate that finite-mass dependence and the meson's internal momentum distribution must be treated consistently when extracting gluon GPDs from exclusive quarkonium production.
By encoding meson structure in a universal DA, the QCD factorization framework enables systematic and unified incorporation of improved inputs from future phenomenological analysis and especially the first-principle lattice QCD calculations. Together with consistent GPD and DA evolution, this framework provides a foundation for precise predictions of exclusive heavy-meson production, which will advance gluon tomography at the future EIC.

\section*{Acknowledgment}

We thank Long-Bin Chen and Bo-Xuan Shi for illuminating discussions. 
This work is partially supported by the Office of Science of the U.S. Department of Energy under Contract No. DE-AC02-05CH11231.
Y.G. is supported by the U.S. Department
of Energy under Contract No. DE-SC0012704, and also by LDRD funds from Brookhaven Science Associates.
Y.J. is grateful for the support in part from the National Natural Science Foundation of China with Grant No.12535006, from the Guangdong Basic and Applied Basic Research Foundation under grant No. 2026A1515011306, and from the University Development Fund of the Chinese University of Hong Kong, Shenzhen, under grant No. UDF01003869. X. J. is partially supported by a grant from National Science Foundation of China, No. 12635004 and Thomas and Linda Lau Family Foundation.
This work is also partially supported by the U.S. Department of Energy, Office of Science, Office of Nuclear Physics, under the umbrella of the Quark-Gluon Tomography (QGT) Topical Collaboration with Award DE-SC0023646.
\bibliographystyle{apsrev4-1}
\bibliography{refs.bib}

\end{document}